\documentstyle[10pt]{article}
\begin{document}
\renewcommand{\baselinestretch}{0.9} \small \normalsize
\begin{center}
{\bf \large {\bf ~~~~~~~~~~~~~~~~~~~~~~\\

~~~~~~~~~~~~~~~~~~~~~~~~~~~~~~~~~\\



DIGRAPH DESCRIPTION OF K-INTERCHANGE TECHNIQUE FOR OPTIMIZATION
OVER PERMUTATIONS AND ADAPTIVE ALGORITHM SYSTEM\\
(Foundation of Computing and Decision Sciences, 26/3, 225-235,
2001)}}
\\[1.0cm]
\end{center}
\begin{center}
Mark Sh. LEVIN*
\\[2.0cm]
\end{center}
~~~~{\bf Abstract.}
The paper describes a general glance to the use of element exchange
techniques for optimization over permutations.
A multi-level description of problems is proposed which is a
fundamental to understand nature and complexity of optimization
problems over permutations
(e.g., ordering, scheduling, traveling salesman problem).
The description is based on permutation neighborhoods of several kinds
(e.g., by improvement of an objective function).
Our proposed operational digraph and
its kinds can be considered as a way to
understand convexity and polynomial solvability
for combinatorial optimization problems
over permutations.
Issues of an analysis of problems and a design of hierarchical
heuristics are discussed.
The discussion leads to a multi-level
adaptive algorithm system which analyzes an individual problem and
selects / designs a solving strategy (trajectory).
\\[1.6cm]
\newcounter{cms}
\setlength{\unitlength}{1mm}
{\bf \large {\bf 1. Introduction}}
\\[0.5cm]
For many years efforts of researchers in combinatorial optimization
were oriented to the design of effective (polynomial)
algorithms for problems on permutations.
Scheduling problems and linear ordering problems
are representatives of the problems over permutations.
In many cases, effective algorithms are based on the use
of local optimization techniques as
{\it two neighbor elements exchange} techniques
which effectively lead to a global optimum.
The following works can be pointed out, for example:
       Adolphson and Hu [1],
       Borie [2],
       Conway et al. [3],
       Hardy et al. [6],
       Johnson [7],
       Levin [8],
       Monma and Sidney [10],
       Sidney [13], and
       Smith [15].

~~~~Based on the work of Smith,
Elmaghraby proposed a graph-theoretical interpretation
for the corresponding \(2\)-search problem
(i.e., interchange of two neighbor elements) [4].

~~~~   This paper describes an extension of the result
of Elmaghraby for \(k\)-search problems.
Here a general multiple level
digraph-description for the optimization problems

\begin{picture}(100,5)
\put(4,00){\makebox(0,0)[bl]
{ *~~Ben-Gurion University, Beer Sheva, Israel; email: mslevin@acm.org}}
\put(00,4){\line(1,0){60}}
\end{picture}
\\[0.3cm]
over permutations on the basis of \(k\)-element exchange
(\(k=n,(n-1),...,2\)) is proposed [14].
The viewpoint herein provides insight which can be incorporated into
the design and analysis of a hierarchical algorithm system.
The system involves a control unit with the following functions:
(a) an analysis of individual problems,
(b) the selection / design of a solving strategy, and
(c) on-line adaptation of the problem solving process.
The described material is a research in progress.
\\[0.7cm]
{\bf \large  {\bf 2. Graph Description}}
\\[0.5cm]
{\bf 2.1. Formulation of Problem Instance}
\\[0.5cm]
In this section, an example is presented as a basis
for our further problem analysis and formulation.
The following problem instance is considered.
There is a set of elements \(S\) and
a function \(f:~S~ \Rightarrow ~R\).
The problem is:
\\[0.15cm]

~~~~{\it Find} ~~\(\min \{f(s) | s \in S\}\).
\\[0.15cm]

~~~~A {\it Neighborhood}  \(A(s_{o}) \subseteq S\)
is associated with \(\forall s \in S\)
such that \(s \in A(s_{o})\).

~~~~{\it Neighborhood Search} is described as follows:
~~Given \(\forall s \in S\),
try to find \(t\in A(s)\) such that \(f(t) < f(s)\).
If no such \(t\) exists, then \(s\) is locally optimal STOP.
Otherwise, replace \(s\) by \(t\) and
repeat until a local optimum is found.

~~~~Given a set \(A=\{A(s)|s\in S\}\) we are particularly interested
in the question of whether neighborhood search using \(A\)
is guaranteed to arrive at local optimum
which is a global optimum as well.

~~~~From now on we will focus upon problems with
\(S= \) {\it the set of permutations of} \(\{1,...,n\}\).
   We will define the \(k-\){\it neighborhood} \(A_{k}(s)\) of
\(s \in S\)  for \(k \leq n-1\)~ to be the set of permutations
which can be obtained from \(s\) by selecting \(k\) adjacent elements
of \(s\) and replacing
them with any permutation of these \(k\) elements
(obviously \(s\in A_{k}(s)\)).
Let \(A(k)=\{A_{k}| s \in S\}\).

~~~~A \(k-\){\it search algorithm}
is a neighborhood search algorithm which uses \(A(k)\) for
its neighborhoods.
   The \(2-\){\it search algorithm} is the basic
{\it adjacent interchange algorithm}.
This method is well-known to be optimal for several sequencing problems
without precedence constraints, e.g.,
   weighted average completion time [15],
   two-machine flow shop [7], etc.
  The following question is posed (Jeffrey B. Sidney):
\\[0.15cm]
~~~~Are there objective functions \(f:~S \Rightarrow R\) for which
the ~\(k-\){\it search algorithm}~ always produces an optimal solution,
but the  ~\((k-1)-\){\it search algorithm}~
does not, for some \(k \geq 3\) ?
\\[0.15cm]
~~~~The answer is YES.

~~~~Let \(n=4\) and set \(f((1,2,3,4))=0\).
Define \(f(s)\) to be the minimum number of neighborhood search steps
that must be executed using neighborhoods \(A(3)\) as
the set of neighborhoods to convert the permutation \(s\)
to the permutation \((1,2,3,4)\).
This function is tabulated in Table 1.
Then the \(3-\){\it search algorithm} is sure to yield the optimum, while
\(2-\){\it search algorithm} will not.

~~~~Consider \(s=(4,3,1,2)\). Then \(f(s)=2\).
Using pairwise interchange (\(2-\){\it search}),
we may reach the following other neighborhood points:

~~~~\((3,4,1,2)\) with \(f(3,4,1,2)=2\),

~~~~\((4,1,3,2)\) with \(f(4,1,3,2)=2\), and

~~~~\((4,3,2,1)\) with \(f(4,3,2,1)=3\).

~~~~Thus, pairwise interchange (\(2-\){\it search})
terminates with the local optimum \((4,3,1,2)\),
while (\(3-\){\it search}) finds the optimum in two iterations of
neighborhood search.

\begin{center}
\begin{picture}(100,66)
\put(23,60){\makebox(0,0)[bl]
{{\bf Table 1. Numerical Example}}}
\put(00,2){\line(00,01){56}}
\put(10,2){\line(00,01){56}}
\put(30,2){\line(00,01){56}}
\put(50,2){\line(00,01){56}}
\put(0,2){\line(01,0){51}}
\put(0,52){\line(01,0){51}}
\put(0,58){\line(01,0){51}}

\put(51,2){\line(00,01){56}}
\put(61,2){\line(00,01){56}}
\put(81,2){\line(00,01){56}}
\put(101,2){\line(00,01){56}}
\put(51,2){\line(01,0){50}}
\put(51,52){\line(01,0){50}}
\put(51,58){\line(01,0){50}}

\put(3,54){\makebox(0,0)[bl]{No}}
\put(20,54){\makebox(0,0)[bl]{\(s\)}}
\put(37,54){\makebox(0,0)[bl]{\(f(s)\)}}
\put(54,54){\makebox(0,0)[bl]{No}}
\put(71,54){\makebox(0,0)[bl]{\(s\)}}
\put(88,54){\makebox(0,0)[bl]{\(f(s)\)}}

\put(4,48){\makebox(0,0)[bl]{{\bf 1}}}
\put(4,44){\makebox(0,0)[bl]{{\bf 2}}}
\put(4,40){\makebox(0,0)[bl]{{\bf 3}}}
\put(4,36){\makebox(0,0)[bl]{{\bf 4}}}
\put(4,32){\makebox(0,0)[bl]{{\bf 5}}}
\put(4,28){\makebox(0,0)[bl]{{\bf 6}}}
\put(4,24){\makebox(0,0)[bl]{{\bf 7}}}
\put(4,20){\makebox(0,0)[bl]{{\bf 8}}}
\put(4,16){\makebox(0,0)[bl]{{\bf 9}}}
\put(4,12){\makebox(0,0)[bl]{{\bf 10}}}
\put(4,08){\makebox(0,0)[bl]{{\bf 11}}}
\put(4,04){\makebox(0,0)[bl]{{\bf 12}}}

\put(54,48){\makebox(0,0)[bl]{{\bf 13}}}
\put(54,44){\makebox(0,0)[bl]{{\bf 14}}}
\put(54,40){\makebox(0,0)[bl]{{\bf 15}}}
\put(54,36){\makebox(0,0)[bl]{{\bf 16}}}
\put(54,32){\makebox(0,0)[bl]{{\bf 17}}}
\put(54,28){\makebox(0,0)[bl]{{\bf 18}}}
\put(54,24){\makebox(0,0)[bl]{{\bf 19}}}
\put(54,20){\makebox(0,0)[bl]{{\bf 20}}}
\put(54,16){\makebox(0,0)[bl]{{\bf 21}}}
\put(54,12){\makebox(0,0)[bl]{{\bf 22}}}
\put(54,08){\makebox(0,0)[bl]{{\bf 23}}}
\put(54,04){\makebox(0,0)[bl]{{\bf 24}}}

\put(68,48){\makebox(0,0)[bl]{\(1234\)}}
\put(68,44){\makebox(0,0)[bl]{\(1243\)}}
\put(68,40){\makebox(0,0)[bl]{\(1324\)}}
\put(68,36){\makebox(0,0)[bl]{\(1342\)}}
\put(68,32){\makebox(0,0)[bl]{\(1423\)}}
\put(68,28){\makebox(0,0)[bl]{\(1432\)}}
\put(68,24){\makebox(0,0)[bl]{\(2134\)}}
\put(68,20){\makebox(0,0)[bl]{\(2143\)}}
\put(68,16){\makebox(0,0)[bl]{\(2314\)}}
\put(68,12){\makebox(0,0)[bl]{\(2341\)}}
\put(68,08){\makebox(0,0)[bl]{\(2413\)}}
\put(68,04){\makebox(0,0)[bl]{\(2431\)}}

\put(17,48){\makebox(0,0)[bl]{\(3124\)}}
\put(17,44){\makebox(0,0)[bl]{\(3142\)}}
\put(17,40){\makebox(0,0)[bl]{\(3214\)}}
\put(17,36){\makebox(0,0)[bl]{\(3241\)}}
\put(17,32){\makebox(0,0)[bl]{\(3412\)}}
\put(17,28){\makebox(0,0)[bl]{\(3421\)}}

\put(17,24){\makebox(0,0)[bl]{\(4123\)}}
\put(17,20){\makebox(0,0)[bl]{\(4132\)}}
\put(17,16){\makebox(0,0)[bl]{\(4213\)}}
\put(17,12){\makebox(0,0)[bl]{\(4231\)}}
\put(17,08){\makebox(0,0)[bl]{\(4312\)}}
\put(17,04){\makebox(0,0)[bl]{\(4321\)}}

\put(91,48){\makebox(0,0)[bl]{\(0\)}}
\put(91,44){\makebox(0,0)[bl]{\(1\)}}
\put(91,40){\makebox(0,0)[bl]{\(1\)}}
\put(91,36){\makebox(0,0)[bl]{\(1\)}}
\put(91,32){\makebox(0,0)[bl]{\(1\)}}
\put(91,28){\makebox(0,0)[bl]{\(1\)}}
\put(91,24){\makebox(0,0)[bl]{\(1\)}}
\put(91,20){\makebox(0,0)[bl]{\(2\)}}
\put(91,16){\makebox(0,0)[bl]{\(1\)}}
\put(91,12){\makebox(0,0)[bl]{\(2\)}}
\put(91,08){\makebox(0,0)[bl]{\(2\)}}
\put(91,04){\makebox(0,0)[bl]{\(2\)}}

\put(40,48){\makebox(0,0)[bl]{\(1\)}}
\put(40,44){\makebox(0,0)[bl]{\(2\)}}
\put(40,40){\makebox(0,0)[bl]{\(1\)}}
\put(40,36){\makebox(0,0)[bl]{\(2\)}}
\put(40,32){\makebox(0,0)[bl]{\(2\)}}
\put(40,28){\makebox(0,0)[bl]{\(2\)}}
\put(40,24){\makebox(0,0)[bl]{\(2\)}}
\put(40,20){\makebox(0,0)[bl]{\(2\)}}
\put(40,16){\makebox(0,0)[bl]{\(2\)}}
\put(40,12){\makebox(0,0)[bl]{\(3\)}}
\put(40,08){\makebox(0,0)[bl]{\(2\)}}
\put(40,04){\makebox(0,0)[bl]{\(3\)}}
\end{picture}
\\[0.7cm]
\end{center}
{\bf 2.2. General Description}
\\[0.5cm]
Here a graph description
of the initial domain of permutations
(an analogue of argument space \(X\)
for a function \(f(x), x \in X \)) is examined.
Let \(G=(P,E)\) be a graph in which
vertex set \(P\) corresponds to permutations
and edge set \(E\) corresponds to a "closeness" of permutation
pairs.
Evidently, some of the edges of \(E\)
can be considered as defining possible element interchanges.

~~~~Thus, we can consider graph of \(k-\){\it closeness}
(from \(k-\){\it interchange} viewpoint of view) as follows:
\(G^{k}=(P,E^{k})\).
By analogy, we get digraph \(D^{k}=(P,O^{k})\) where
the following conditions hold:

~~~~   (1) ~\(p_{i}, p_{j} \in P\), ~\(p_{i},p_{j} \in E\),
       and \(p_{i}\) and \(p_{j}\) are "close";

~~~~   (2) ~\((p_{i},p_{j}) \in O^{k}\) if and only if \(p_{j}\)
        can be obtained from \(p_{i}\) by \(k-\)interchange.

~~~~ We can examine a multi-level description of
the above-mentioned operation set \(O^{k}\)
(\(O^{k} \subseteq E^{k}\)):
\(k\) corresponds to possible \(k-\){\it element} interchange.
Thus we arrive at the following possibilities:
\\[0.15cm]

~~~~ \(n-\){\it exchange} algorithm on a graph \(G=(P,E)\):
every permutation is adjacent to every other one,

~~~~ \((n-k)-\){\it exchange} algorithm on digraph
\(D^{n-k}=(P,O^{n-k})\) for \(1 \leq k \leq n-k\).

~~~~.~.~.
\\[0.15cm]

~~~~The \(2-\){\it interchange} (adjacent interchange) algorithm
uses the digraph \(D^{2}=(P,O^{2})\).
A related generalized description for the traveling salesman problem
has been described by Reinelt in [11].
The following result is obvious:
\[O^{k} \subseteq O^{k+1}, ~\forall k = 2,...,n-1.\]
Two properties of interest, which may or may not hold for given
problem, are:
\\[0.15cm]

~~~~{\bf Property 1.} \(\forall p \in P\) exists a path in
\(D\) which leads to an optimal permutation.
\\[0.15cm]

~~~~{\bf Property 2.} \(\forall p \in P\) exists a path in
\(D\) for which the following hold:

~~~~(1) the path leads to an optimal point;

~~~~(2) the length of the path which corresponds to the
number of interchanges is polynomial in \(n\) (steps of interchanges).
\\[0.15cm]

~~~~It is evident that Property 2 implies Property 1.

~~~~Now it is reasonable to consider the following observations:

~~~~{\bf 1.} The structure of the digraph \(D^{k}\) (i.e., its connectivity)
for a certain kind
of problem defines its complexity, e.g., the existence and length
(polynomial in \(n\) or not) of the shortest path from a point to the
optimal.

~~~~{\bf 2.} Not all optimization problems on permutations have
connected digraph \(D^{k}\).

~~~~{\bf 3.} Known combinatorial problems
for which polynomial \(2-\){\it interchange} algorithms
exist have connected digraph \(D^{2}\) with very "good"
structure (e.g., tree).

~~~~{\bf 4.} For "hard" combinatorial problems
the digraph \(D^{k}\) is unconnected at small levels of \(k\).
In other words, only the use of \(k-\){\it interchange algorithm}
for higher \(k\), perhaps even \(n\) will guarantee reaching the optimum.

~~~~{\bf 5.} A digraph \(D^{k}\) may correspond to a problem
with more than one path to an optimal point(s).
\\[0.7cm]
{\bf \large {\bf 3. Neighborhoods and Operational Digraph}}
\\[0.5cm]
In section 2, neighborhood \(A_{k}(s)\) for element \(s \in S\)
was defined.
Now we examine a function \(f(x)\) where
\(x=(x_{1},...,x_{i},...,x_{n})\) is a permutation.
It is assumed that \(f(x)\) is integer-valued.
We define specific types of neighborhoods as follows:
\\[0.15cm]

~~~~{\bf Definition.}

~~~~Let \(V^{k}(s)\) be a
\(k-\){\it interchange} neighborhood
of point \(s\)  defined  by \(x \in V^{k}(s)\)
if and only if \(x \neq s\) and \(x\) can be obtained from
\(s\) via a single \(k-\){\it interchange}.

~~~~Let \(V^{k<}(s) \subseteq V^{k}(s)\)
be that subset of \(V^{k}(s)\)
such that  \(f(x) < f(s)\).

~~~~Let \(V^{k<=}(s) \subseteq V^{k}(s)\)
be that subset of \(V^{k}(s)\) such that \(f(x) \leq f(s)\).
\\[0.15cm]

~~~~\(D^{k<}\) and \(D^{k<=}\) are digraphs which correspond to
~\(k-\){\it interchange} algorithms on the basis of
{\it improvement} of \(f(x)\)
and {\it improvement} or {\it equivalence} of \(f(x)\), respectively.
Note in the case of {\it equivalence} each {\it equivalence}-edge
in \(D^{k<=}\) will correspond to two arcs with opposite directions.

~~~~As a result of the definition above we obtain the following:
\[O^{k<} \subseteq O^{k<=}, ~V^{k-1} \subseteq V^{k},
~V^{k<} \subseteq V^{k<=}.\]

~~~~Now let us examine numerical examples
based upon the function shown in Table 1.
Fig. 1. shows the digraph \(D^{2<} = (P,O^{k}) \)
where \((p_{i},p_{j}) \in O^{k}\)
if and only if
 \(p_{j} \in V^{2<} (p_{i}) \), i.e.,
\(p_{j}\)
can be obtained from
\(p_{i}\)
by adjacent interchange, and
\( f(p_{j}) < f(p_{i})\).
Note that the graph is not connected, and in fact the optimal point
\((1,2,3,4)\) is not connected to and can be reached by
\(2-\){\it interchange} from only four other points, mainly
  ~(\(2,1,4,3\)),
  ~(\(1,2,4,3\)),
  ~(\(1,3,2,4\)),
   and
  ~(\(2,1,3,4\)).

\begin{center}
\begin{picture}(115,118)
\put(15,05){\makebox(0,0)[bl]
{{\bf Fig. 1. Illustration for digraph \(D^{2<}\)}}}

\put(2,110){\makebox(0,0)[bl]{\(f(x)=3\)}}
\put(32,110){\makebox(0,0)[bl]{\(f(x)=2\)}}
\put(62,110){\makebox(0,0)[bl]{\(f(x)=1\)}}
\put(92,110){\makebox(0,0)[bl]{\(f(x)=0\)}}

\put(8,55){\oval(16,6)}
\put(2,54){\makebox(0,0)[bl]{{\bf 10:}\(4231\)}}
\put(8,47){\oval(16,6)}
\put(2,46){\makebox(0,0)[bl]{{\bf 12:}\(4321\)}}

\put(16,47){\vector(1,0){14}}
\put(16,55){\vector(1,0){14}}
\put(16,47){\line(1,2){7}}
\put(23,61){\line(0,1){18}}
\put(23,79){\vector(1,0){7}}
\put(16,55){\line(1,-1){4}}
\put(20,51){\line(0,-1){36}}
\put(20,15){\vector(1,0){10}}
\put(38,103){\oval(16,6)}
\put(32,102){\makebox(0,0)[bl]{{\bf 2~:}\(3142\)}}
\put(38,95){\oval(16,6)}
\put(32,94){\makebox(0,0)[bl]{{\bf 4~:}\(3241\)}}
\put(38,87){\oval(16,6)}
\put(32,86){\makebox(0,0)[bl]{{\bf 5~:}\(3412\)}}
\put(38,79){\oval(16,6)}
\put(32,78){\makebox(0,0)[bl]{{\bf 6~:}\(3421\)}}
\put(38,71){\oval(16,6)}
\put(32,70){\makebox(0,0)[bl]{{\bf 7~:}\(4123\)}}
\put(38,63){\oval(16,6)}
\put(32,62){\makebox(0,0)[bl]{{\bf 8~:}\(4132\)}}
\put(38,55){\oval(16,6)}
\put(32,54){\makebox(0,0)[bl]{{\bf 9~:}\(4213\)}}
\put(38,47){\oval(16,6)}
\put(32,46){\makebox(0,0)[bl]{{\bf 11:}\(4312\)}}
\put(38,39){\oval(16,6)}
\put(32,38){\makebox(0,0)[bl]{{\bf 20:}\(2143\)}}
\put(38,31){\oval(16,6)}
\put(32,30){\makebox(0,0)[bl]{{\bf 22:}\(2341\)}}
\put(38,23){\oval(16,6)}
\put(32,22){\makebox(0,0)[bl]{{\bf 23:}\(2413\)}}
\put(38,15){\oval(16,6)}
\put(32,14){\makebox(0,0)[bl]{{\bf 24:}\(2431\)}}
\put(46,103){\line(1,0){7}}
\put(53,103){\vector(1,-1){7}}

\put(46,95){\line(1,0){7}}
\put(53,95){\vector(1,-1){7}}

\put(46,39){\vector(1,0){14}}
\put(46,31){\vector(1,0){14}}
\put(46,71){\vector(1,-1){14}}
\put(46,63){\vector(1,-1){14}}

\put(46,103){\line(1,-1){6}}
\put(52,97){\line(0,-1){18}}
\put(52,79){\vector(1,-2){8}}
\put(46,39){\line(1,1){9}}
\put(55,48){\line(0,1){26}}
\put(55,74){\vector(1,1){5}}

\put(68,95){\oval(16,6)}
\put(62,94){\makebox(0,0)[bl]{{\bf 1~:}\(3124\)}}
\put(68,87){\oval(16,6)}
\put(62,86){\makebox(0,0)[bl]{{\bf 3~:}\(3214\)}}
\put(68,79){\oval(16,6)}
\put(62,78){\makebox(0,0)[bl]{{\bf 14:}\(1243\)}}
\put(68,71){\oval(16,6)}
\put(62,70){\makebox(0,0)[bl]{{\bf 15:}\(1324\)}}
\put(68,63){\oval(16,6)}
\put(62,62){\makebox(0,0)[bl]{{\bf 16:}\(1342\)}}
\put(68,55){\oval(16,6)}
\put(62,54){\makebox(0,0)[bl]{{\bf 17:}\(1423\)}}
\put(68,47){\oval(16,6)}
\put(62,46){\makebox(0,0)[bl]{{\bf 18:}\(1432\)}}
\put(68,39){\oval(16,6)}
\put(62,38){\makebox(0,0)[bl]{{\bf 19:}\(2134\)}}
\put(68,31){\oval(16,6)}
\put(62,30){\makebox(0,0)[bl]{{\bf 21:}\(2314\)}}
\put(98,39){\oval(16,6)}
\put(92,38){\makebox(0,0)[bl]{{\bf 13:}\(1234\)}}

\put(76,39){\vector(1,0){14}}
\put(76,79){\line(1,0){9}}
\put(85,79){\line(0,-1){30}}
\put(85,49){\vector(1,-2){5}}
\put(76,71){\line(1,0){8}}
\put(84,71){\line(0,-1){24}}
\put(84,47){\vector(2,-3){5.5}}
\end{picture}
\end{center}

~~~~Fig. 2 depicts \(D^{2<=}\).
In this case, there exists a path
from each permutation \(\forall x \in X\) to the optimum point.
  Fig. 3 demonstrates a simple procedure for finding the
\(3-\){\it neighborhood} of the permutation \( (1,2,3,4) \).
Every permutation of every contiguous set of length \(3\)
in \((1,2,3,4)\) is listed, and duplicates are crossed out.
Note that the digraph \(D^{3<}\) (\(3-\){\it interchange} algorithm)
includes a path from every point to the optimal point.

~~~~Furthermore, the graphical structure of our problem can be analyzed.
Without loss of generality we specify
\( (1,...,n) \) as an optimal permutation, and, in a similar fashion to
section 2, define for \(2 \leq k \leq n\) the functions
~\(g_{k}(s): P \rightarrow R\)~ to be the minimal number of
neighborhood search steps needed to transform a permutation \(s\)
into \(1,...n\).
The graph \(D^{k<}\) is defined as before.
Let \(L = \max \{g_{k}(s)|s \in P\}\).
\(L + 1\) represents the number of levels in \(D^{k<}\)
where level \(i\) is defined to be the set
\(\{s|g_{k}=i\}\).
It is clear from the above definition that \(D^{k<}\)
is a connected digraph, and that there is a directed path from any
permutation to \( (1,...,n) \).

\begin{center}
\begin{picture}(115,117)
\put(15,05){\makebox(0,0)[bl]
{{\bf Fig. 2. Illustration for digraph \(D^{2<=}\)}}}

\put(2,110){\makebox(0,0)[bl]{\(f(x)=3\)}}
\put(32,110){\makebox(0,0)[bl]{\(f(x)=2\)}}
\put(62,110){\makebox(0,0)[bl]{\(f(x)=1\)}}
\put(92,110){\makebox(0,0)[bl]{\(f(x)=0\)}}

\put(8,55){\oval(16,6)}
\put(2,54){\makebox(0,0)[bl]{{\bf 10:}\(4231\)}}
\put(8,47){\oval(16,6)}
\put(2,46){\makebox(0,0)[bl]{{\bf 12:}\(4321\)}}
\put(16,55){\line(1,-1){4}}
\put(16,47){\line(1,1){4}}

\put(16,47){\vector(1,0){14}}
\put(16,55){\vector(1,0){14}}
\put(16,47){\line(1,2){7}}
\put(23,61){\line(0,1){18}}
\put(23,79){\vector(1,0){7}}
\put(16,55){\line(2,-1){7}}
\put(23,51.5){\line(0,-1){36.5}}
\put(23,15){\vector(1,0){7}}
\put(38,103){\oval(16,6)}
\put(32,102){\makebox(0,0)[bl]{{\bf 2~:}\(3142\)}}
\put(38,95){\oval(16,6)}
\put(32,94){\makebox(0,0)[bl]{{\bf 4~:}\(3241\)}}
\put(38,87){\oval(16,6)}
\put(32,86){\makebox(0,0)[bl]{{\bf 5~:}\(3412\)}}
\put(38,79){\oval(16,6)}
\put(32,78){\makebox(0,0)[bl]{{\bf 6~:}\(3421\)}}
\put(38,71){\oval(16,6)}
\put(32,70){\makebox(0,0)[bl]{{\bf 7~:}\(4123\)}}
\put(38,63){\oval(16,6)}
\put(32,62){\makebox(0,0)[bl]{{\bf 8~:}\(4132\)}}
\put(38,55){\oval(16,6)}
\put(32,54){\makebox(0,0)[bl]{{\bf 9~:}\(4213\)}}
\put(38,47){\oval(16,6)}
\put(32,46){\makebox(0,0)[bl]{{\bf 11:}\(4312\)}}
\put(38,39){\oval(16,6)}
\put(32,38){\makebox(0,0)[bl]{{\bf 20:}\(2143\)}}
\put(38,31){\oval(16,6)}
\put(32,30){\makebox(0,0)[bl]{{\bf 22:}\(2341\)}}
\put(38,23){\oval(16,6)}
\put(32,22){\makebox(0,0)[bl]{{\bf 23:}\(2413\)}}
\put(38,15){\oval(16,6)}
\put(32,14){\makebox(0,0)[bl]{{\bf 24:}\(2431\)}}

\put(46,95){\line(1,-1){5}}
\put(51,90){\line(0,-1){54}}
\put(46,31){\line(1,1){5}}

\put(46,55){\line(1,0){6}}
\put(52,55){\line(0,-1){32}}
\put(46,23){\line(1,0){6}}

\put(46,87){\line(1,0){4}}
\put(50,87){\line(0,-1){40}}
\put(46,47){\line(1,0){4}}

\put(30,87){\line(-1,-1){4}}
\put(30,79){\line(-1,1){4}}

\put(30,71){\line(-1,-1){4}}
\put(30,63){\line(-1,1){4}}

\put(30,39){\line(-1,0){6}}
\put(24,39){\line(0,-1){16}}
\put(30,23){\line(-1,0){6}}

\put(30,103){\line(-1,0){5}}
\put(25,103){\line(0,-1){16}}
\put(30,87){\line(-1,0){5}}

\put(30,95){\line(-1,0){7}}
\put(23,95){\line(0,-1){12.5}}
\put(30,79){\line(-2,1){7}}

\put(30,71){\line(-1,0){5}}
\put(25,71){\line(0,-1){11}}
\put(30,55){\line(-1,1){5}}

\put(30,63){\line(-1,0){4}}
\put(26,63){\line(0,-1){12}}
\put(30,47){\line(-1,1){4}}

\put(30,31){\line(-1,0){4}}
\put(26,31){\line(0,-1){12}}
\put(30,15){\line(-1,1){4}}

\put(46,103){\line(1,0){7}}
\put(53,103){\vector(1,-1){7}}

\put(46,95){\line(1,0){7}}
\put(53,95){\line(0,-1){8}}
\put(53,87){\vector(1,0){7}}

\put(46,39){\vector(1,0){14}}
\put(46,31){\vector(1,0){14}}
\put(46,71){\vector(1,-1){14}}
\put(46,63){\vector(1,-1){14}}

\put(46,103){\line(1,-1){6}}
\put(52,97){\line(0,-1){18}}
\put(52,79){\vector(1,-2){8}}
\put(46,39){\line(1,1){9}}
\put(55,48){\line(0,1){26}}
\put(55,74){\vector(1,1){5}}

\put(68,95){\oval(16,6)}
\put(62,94){\makebox(0,0)[bl]{{\bf 1~:}\(3124\)}}
\put(68,87){\oval(16,6)}
\put(62,86){\makebox(0,0)[bl]{{\bf 3~:}\(3214\)}}
\put(68,79){\oval(16,6)}
\put(62,78){\makebox(0,0)[bl]{{\bf 14:}\(1243\)}}
\put(68,71){\oval(16,6)}
\put(62,70){\makebox(0,0)[bl]{{\bf 15:}\(1324\)}}
\put(68,63){\oval(16,6)}
\put(62,62){\makebox(0,0)[bl]{{\bf 16:}\(1342\)}}
\put(68,55){\oval(16,6)}
\put(62,54){\makebox(0,0)[bl]{{\bf 17:}\(1423\)}}
\put(68,47){\oval(16,6)}
\put(62,46){\makebox(0,0)[bl]{{\bf 18:}\(1432\)}}
\put(68,39){\oval(16,6)}
\put(62,38){\makebox(0,0)[bl]{{\bf 19:}\(2134\)}}
\put(68,31){\oval(16,6)}
\put(62,30){\makebox(0,0)[bl]{{\bf 21:}\(2314\)}}
\put(98,39){\oval(16,6)}
\put(92,38){\makebox(0,0)[bl]{{\bf 13:}\(1234\)}}

\put(76,87){\line(1,-1){6}}
\put(82,81){\line(0,-1){44}}
\put(76,31){\line(1,1){6}}

\put(76,95){\line(1,-1){5}}
\put(81,90){\line(0,-1){14}}
\put(76,71){\line(1,1){5}}

\put(76,79){\line(1,-1){5}}
\put(81,74){\line(0,-1){14}}
\put(76,55){\line(1,1){5}}

\put(76,63){\line(1,-1){4}}
\put(80,59){\line(0,-1){8}}
\put(76,47){\line(1,1){4}}

\put(60,95){\line(-1,-1){4}}
\put(60,87){\line(-1,1){4}}

\put(60,39){\line(-1,-1){4}}
\put(60,31){\line(-1,1){4}}

\put(60,55){\line(-1,-1){4}}
\put(60,47){\line(-1,1){4}}

\put(76,39){\vector(1,0){14}}
\put(76,79){\line(1,0){9}}
\put(85,79){\line(0,-1){30}}
\put(85,49){\vector(1,-2){5}}
\put(76,71){\line(1,0){8}}
\put(84,71){\line(0,-1){24}}
\put(84,47){\vector(2,-3){5.5}}
\end{picture}
\\[8pt]
\end{center}
\begin{center}
\begin{picture}(90,57)
\put(0,04){\makebox(0,0)[bl]
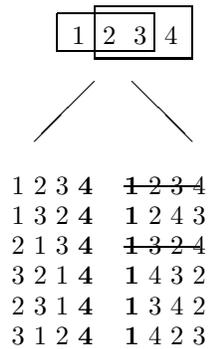
{{\bf Fig. 3. Illustration for 3-neighborhood}}}

\put(31,45){\line(-1,-1){8}}
\put(36,45){\line(1,-1){8}}

\put(31,48){\line(0,1){7}}
\put(44,48){\line(0,1){7}}
\put(31,55){\line(1,0){13}}
\put(31,48){\line(1,0){13}}

\put(26,49){\line(0,1){5}}
\put(39,49){\line(0,1){5}}
\put(26,54){\line(1,0){13}}
\put(26,49){\line(1,0){13}}
\put(28,50){\makebox(0,0)[bl]{\(1~~2~~3~~4\)}}

\put(35,31){\line(1,0){10}}
\put(35,23){\line(1,0){10}}

\put(35,30){\makebox(0,0)[bl]{{\bf 1}~2~3~4}}
\put(35,26){\makebox(0,0)[bl]{{\bf 1}~2~4~3}}
\put(35,22){\makebox(0,0)[bl]{{\bf 1}~3~2~4}}
\put(35,18){\makebox(0,0)[bl]{{\bf 1}~4~3~2}}
\put(35,14){\makebox(0,0)[bl]{{\bf 1}~3~4~2}}
\put(35,10){\makebox(0,0)[bl]{{\bf 1}~4~2~3}}

\put(20,30){\makebox(0,0)[bl]{1~2~3~{\bf 4}}}
\put(20,26){\makebox(0,0)[bl]{1~3~2~{\bf 4}}}
\put(20,22){\makebox(0,0)[bl]{2~1~3~{\bf 4}}}
\put(20,18){\makebox(0,0)[bl]{3~2~1~{\bf 4}}}
\put(20,14){\makebox(0,0)[bl]{2~3~1~{\bf 4}}}
\put(20,10){\makebox(0,0)[bl]{3~1~2~{\bf 4}}}

\end{picture}
\end{center}

~~~~Now let ~\(f : P \rightarrow R\)~ represents a function to be
minimized, and without loss of generality let
\(s=(1,...n)\) be a permutation which minimizes \(f\).
Define \(V^{k<}\)  and \(V^{k<=}\) based upon the function \(f\).
If \(t\) is a local optimum, it is follows that
\(|V^{k<}(t)|=0\).
However, \(V^{k<=}(t)\) may be non-empty in the case of ties, and
this applies to \(V^{k<=}(s)\) as well.
\\[0.7cm]
{\bf \large {\bf 4. Algorithm System}}
\\[0.5cm]
{\bf 4.1. Implementation Issues}
\\[0.5cm]
Now let us consider two basic implementation issues .
Several observations are in order:

~~~~{\it 1.} If for every non-optimal \(x\), \(|V^{k<}| > 0\)
then there exists a path in \(D^{k<}\) from every \(x\) to an optimal point.

~~~~{\it 2.} A necessary and sufficient condition for there to be a
directed path in \(D^{k<=}\) from every non-optimal permutation is the
following:

~~~~For all non-optimal \(x\) either \(|V^{k<}>0|\) or there exists a sequence
\((x=y_{1},...,y_{h})\) such that for \(1 \leq j \leq h-1\) the relationship
\(y_{j=1} \in V^{k<=}(y_{j})\)
and also
\(|V^{k<}| \neq 0\).

~~~~Note that
 the number of levels in the graph \(D^{2}\) is \(n(n-1)/2 + 1\)
and the number of levels is less for \(D^{k}\) with \(k > 2\).

~~~~Now using (i) and (ii) we get:

~~~~{\it 3.} The \(k-\){\it interchange} algorithm yields an optimal
in polynomial time if

~~~~(a)~ \(\forall x ~~|V^{k<}(x)| > 0\) ~and

~~~~(b)~ \(|V^{k<}(x)|\)~ is polynomial in \(n\).

\begin{center}
\begin{picture}(115,60)
\put(17,02){\makebox(0,0)[bl]
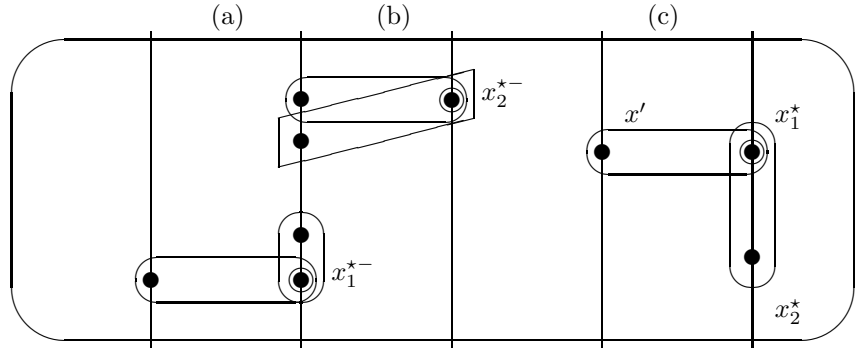
{{\bf Fig. 4. Illustration for local and global optimum}}}

\put(28,52){\makebox(0,0)[bl]{(a)}}
\put(50,52){\makebox(0,0)[bl]{(b)}}
\put(86,52){\makebox(0,0)[bl]{(c)}}

\put(57.5,30){\oval(112,40)}
\put(20,9){\line(0,1){42}}
\put(40,9){\line(0,1){42}}
\put(60,9){\line(0,1){42}}
\put(80,9){\line(0,1){42}}
\put(100,9){\line(0,1){42}}

\put(80,35){\circle*{2}}
\put(90,35){\oval(24,6)}
\put(100,35){\circle*{2}}
\put(100,35){\circle{3}}

\put(100,21){\circle*{2}}
\put(100,21){\circle{1}}
\put(100,28){\oval(6,22)}

\put(103,39){\makebox(0,0)[bl]{\(x_{1}^{\star }\)}}
\put(103,13){\makebox(0,0)[bl]{\(x_{2}^{\star}\)}}
\put(83,39){\makebox(0,0)[bl]{\(x'\)}}

\put(63,39.5){\line(0,1){6.5}}
\put(37,33){\line(0,1){6.5}}
\put(37,39.5){\line(4,1){26}}
\put(37,33){\line(4,1){26}}

\put(40,42){\circle*{2}}
\put(50,42){\oval(24,6)}
\put(60,42){\circle{3}}
\put(60,42){\circle*{2}}

\put(40,36.5){\circle*{2}}
\put(64,42){\makebox(0,0)[bl]{\(x_{2}^{\star -}\)}}

\put(40,21){\oval(6,12)}
\put(40,24){\circle*{2}}
\put(20,18){\circle*{2}}
\put(30,18){\oval(24,6)}
\put(40,18){\circle{3}}
\put(40,18){\circle*{2}}
\put(44,18){\makebox(0,0)[bl]{\(x_{1}^{\star -}\)}}
\end{picture}
\end{center}

~~~~There are many approaches to using \(k-\){\it interchange}.
The choice of the initial point, and,
  where choice exists,
  the choice of next point,
are crucial parts of such algorithm.
In line with many modern optimization approaches,
probabilistic methods may be in order.
Another key issue is identification of the
optimum when it is found. Such identification depends upon the nature
of the function being optimized.
It may be useful also to start with small \(k\) (say \(2\))
and only increase \(k\) to \(k+1\) when a local (but not global)
optimum is reached with \(k-\){\it interchange}.
After an improvement, the algorithm could return to using \(k=2\).
Fig. 4 illustrates possible situations with local and global optimum.
Fig. 5 and 6 depict paths (strategies) to a global optimum for
\(D^{k<}\) and \(D^{k<=}\).

\begin{center}
\begin{picture}(115,56)
\put(23,02){\makebox(0,0)[bl]
{{\bf Fig. 5. Illustration for path \(l(x)\) \(D^{k<}\)}}}

\put(57.5,30){\oval(112,40)}

\put(20,9){\line(0,1){42}}
\put(40,9){\line(0,1){42}}
\put(60,9){\line(0,1){42}}
\put(80,9){\line(0,1){42}}
\put(100,9){\line(0,1){42}}

\put(80,35){\circle*{2}}
\put(90,35){\oval(24,6)}
\put(100,35){\circle*{2}}
\put(100,35){\circle{3}}
\put(60,33){\circle*{2}}
\put(70,33){\oval(24,6)}
\put(80,31){\circle{1}}
\put(80,35){\circle*{2}}

\put(37,22){\line(0,1){7}}
\put(37,22){\line(1,0){26}}
\put(37,29){\line(3,2){26}}
\put(63,22){\line(0,1){24.2}}
\put(60,28){\circle{1}}
\put(60,33){\circle*{2}}
\put(60,38){\circle{1}}
\put(60,42){\circle{1}}

\put(20,22){\circle*{2}}
\put(30,22){\oval(24,6)}
\put(40,20){\circle{1}}
\put(40,24){\circle*{2}}
\end{picture}
\end{center}

\begin{center}
\begin{picture}(115,56)
\put(23,02){\makebox(0,0)[bl]
{{\bf Fig. 6. Illustration for path \(l(x)\) \(D^{k<=}\)}}}
\put(57.5,30){\oval(112,40)}

\put(20,9){\line(0,1){42}}
\put(40,9){\line(0,1){42}}
\put(60,9){\line(0,1){42}}
\put(80,9){\line(0,1){42}}
\put(100,9){\line(0,1){42}}

\put(80,25){\circle*{2}}
\put(90,25){\oval(24,6)}
\put(100,25){\circle{3}}
\put(100,25){\circle*{2}}

\put(80,35){\circle*{2}}
\put(80,30){\oval(6,15)}
\put(80,30){\circle{1}}
\put(80,25){\circle*{2}}

\put(60,33){\circle*{2}}
\put(70,33){\oval(24,9)}
\put(80,35){\circle*{2}}

\put(37,22){\line(0,1){7}}
\put(37,22){\line(1,0){26}}
\put(37,29){\line(3,2){26}}
\put(63,22){\line(0,1){24.2}}

\put(60,28){\circle{1}}
\put(60,33){\circle*{2}}
\put(60,38){\circle{1}}
\put(60,42){\circle{1}}

\put(40,44){\circle*{2}}
\put(40,32){\oval(6,29)}
\put(40,20){\circle{1}}
\put(40,24){\circle*{2}}

\put(20,42){\circle*{2}}
\put(30,42){\oval(24,6)}
\put(40,40){\circle{1}}

\put(40,44){\circle*{2}}
\end{picture}
\end{center}

{\bf 4.2. Solving Trajectory}
\\[0.5cm]
First, the following three kinds of strategy steps exist:
(a) {\it forward} as an improvement;
(b) {\it aside}; and
(c) {\it backward}.
Thus we can consider three one-line trajectories as follows:

~~~~ (1) kind {\bf F} as {\it forward} steps;

~~~~ (2) kind {\bf FA} as {\it forward} and {\it aside} steps; and

~~~~ (3) kind {\bf FAB} as
{\it forward}, {\it aside}, and {\it backward} steps.

~~~~ In addition, it is reasonable to consider multi-line trajectories
which consist of several one-line trajectories:

~~~~  (1) kind {\bf nF} as several trajectories of kind {\bf F};

~~~~  (2) kind {\bf nFA} as several trajectories of kinds {\bf FA};

~~~~  (3) kind {\bf nFAB} as several trajectories of kinds {\bf FAB}.
\\[0.7cm]
{\bf 4.3. Space of Algorithm Control}
\\[0.5cm]
It is reasonable to study properties of an individual problem.
In this case, the following is crucial:

~~~~(a) choice of an initial point (or a set of initial points);

~~~~(b) for a current point \(x\) selection of a next path step
     (i.e., a point \(y\) in neighborhood of \(x\)) because
     selected \(y\) has to lead to an optimal point;

~~~~(c) design a composite strategy or trajectory (composite path);

~~~~(d) on-line analysis of the solved problem and change
      (adaptation) of solving strategy (trajectory) by the following ways:
          (i) examination of new initial point(s);
          (ii) change of strategy types,
          (iii) change of algorithm types
                (decreasing or increasing \(k\)).

~~~~An initial information which is a basis to the problem analysis
consists in types of points (permutations) and their neighborhoods
from viewpoint of a quality of neighbor elements:
~(1) basic cases:
    (i) improvement
          (improvement is possible by moving to a neighbor element),
    (ii) equivalence,
    (iii) optimum; and
~(2) composite cases.
\\[0.7cm]
{\bf 4.4. Structure of Algorithm System}
\\[0.5cm]
We consider
the following three-level structure of the algorithm system [9]:

~~~~{\bf 1.}  Control unit (planning, adaptation):
  (a) selection of algorithms;
  (b) selection / design of composite solving strategy.

~~~~{\bf 2.} Level of execution:
 (a) analysis of an individual ordering problem;
 (b) executing some steps of solving process; and
 (c) analysis of obtained results.

~~~~{\bf 3.} Level of bases / repositories:
  (a) problems and examples;
  (b) base of \(k-\)exchange algorithms \(k=1,...,n\); and
  (c) base of solving strategies.

~~~~The structure is oriented to concepts:
  (i) problem (analysis, approximation),
  (ii) model (selection, prediction),
  (iii) algorithm (selection), and
  (iv) solving strategy (selection, design).
\\[0.7cm]
{\bf \large {\bf 5. Conclusion}}
\\[0.5cm]
Our digraph description of \(k-\){\it interchange} approach is
a good fundamental to analyze types of optimization ordering problems.
The kinds of the proposed operational digraphs correspond to
a problem property that is an analogue of convexity
in continuous optimization.
The considered adaptive algorithm system is close
to optimization method macro-structures
which are applied in global continuous optimization.
Some possible future topics for investigation
are the following:

~~~~{\it 1.}
Studies of well-known combinatorial problems including scheduling problems
on the basis of the proposed approach.

~~~~{\it 2.} Design of a special solving environment and execution of
computing experiments for some well-known ordering problems.

~~~~{\it 3.} Development of probabilistic analysis methods upon the approach.

~~~~{\it 4.} Examination of sensitivity and/or stability for discrete
problems on the basis of our approach.
Problems of graph stability ([5], [12])
are a fundamental for this initiative.

~~~~{\it 5.} Logical functions:
Our description of scheduling problems leads to
  a special class of multi-valued logic functions
  when arguments (\(n\)-size vector) correspond to permutations
  and an ordinal scale is used for the objective function.
  Thus, a scheduling problem can be reformulated as optimization of
  a certain logical function.
  The study of properties for the logical functions is
  of interest (e.g., monotonicity).

~~~~{\it 6.} Examining the possibility of effective algorithm design for
problems where the probability of the existence of paths from each
point to an optimal point is high, although not 100 per cent.

~~~~{\it 7.} Investigation of multicriteria ordering problems.

~~~~{\it 8.} Usage of artificial intelligence approaches
(e.g., anytime algorithms [16]) for the problem analysis
and monitoring the solving process.
\\[0.7cm]
{\bf \large {\bf 6. Acknowledgments}}
\\[0.5cm]
The author acknowledges Jeffrey B. Sidney (University of Ottawa, Canada)
who is the initiator of the research direction [14].
The author thanks the anonymous referee
for careful consideration of the paper and useful comments.
\\[0.7cm]
{\bf \large {\bf References}}
\\[0.5cm]
[1]~~Adolphson D. and Hu T.C.,
 Optimal linear ordering, {\it SIAM J. Appl. Math.},
 {\bf 25}, 1973, 403-423.\\[5pt]
[2]~~Borie R.B.,
 Generation of polynomial-time algorithms for some optimization problems
 on tree-decomposable graphs.
 {\it Algorithmica}, {\bf 14}, 2, 1995, 123-137.\\[5pt]
[3]~~Conway R.W., Maxwell W.L., and Miller L.W.,
 {\it Theory of Scheduling}, Addison-Wesley, Readings, Mass.,
 1967.\\[5pt]
[4]~~Elmaghraby S.E.,
  A graph theoretic interpretation of the sufficiency
  condition for the contiguous binary-switching (CBS)-rule,
  {\it Naval Research Logistics Quarterly},
  {\bf 18}, 3, 1971, 339-344.\\[5pt]
[5]~~Harary F., Norman R.Z., and Carthwright D.,
 {\it Structural Models: An Introduction to the Theory of Directed Graphs},
 Wiley \& Sons, New York, 1965.\\[5pt]
[6]~~Hardy G.H., Littlewood J.E., and Polya G.,
 {\it Inequalities}, 2nd ed., Cambridge University Press,
  Cambridge, 1952.\\[5pt]
[7]~~Johnson S.M.,
  Optimal two- and three-stage
  production schedules with setup times,
  {\it Naval Research Logistic Quarterly},
  {\bf 1}, 1954, 61-67.\\[5pt]
[8]~~Levin M.Sh.,
 Effective solution of certain problems of theory of schedulings of nets,
 {\it Cybernetics (translated from Kibernetika)},
 {\bf 16}, 1-6, 1980, 148-153.\\[5pt]
[9]~~Levin M.Sh.,
  Algorithm systems for combinatorial optimization:
  hierarchical multistage framework, in:
 {\it Proc. of 13th Intl. Conf. on Systems Engineering, Las Vegas,}
 1999, CS109-114.\\[5pt]
[10]~~Monma C.L. and Sidney J.B.,
 Sequencing with series-parallel precedence constraints.
 {\it Mathematics of Operations Research,}
 {\bf 4}, 1979, 215-224.\\[5pt]
[11]~~Reinelt G.,
 {\it The Traveling Salesman, LNCS,} Vol. 840,
 Springer-Verlag, Berlin, 1994.\\[5pt]
[12]~~Roberts F.R.,
 {\it  Discrete Mathematical Models with Applications to Social,
  Biological and Environmental Problems},
  Prentice Hall, NJ, 1976.\\[5pt]
[13]~~Sidney J.B.,
 Decomposition algorithm for single-machine sequencing with precedence
 relations and deferral costs,
 {\it Operations Research}, {\bf 23}, 1975, 283-298.\\[5pt]
[14]~~Sidney J.B. and Levin M.Sh.,
 {\it K-Exchange Technique for Optimization Over
 Permutations}, Working Paper No. 98-04, Faculty of Business
 Administration, Univ. of Ottawa, 1998.\\[5pt]
[15]~~Smith W.E.,
 Various optimizers for single-stage production,
 {\it Nav. Res. Log. Quart.},
  {\bf 3}, 1956, 59-66.\\[5pt]
[16]~~Zilberstein S.,
 Using anytime algorithms in intelligent systems,
 {\it AI magazine}, {\bf 17}, 3, 1996, 73-83.\\[5pt]

 \end{document}